\documentclass[11pt]{article}
\usepackage{amssymb,amsmath,latexsym}
\usepackage{mathrsfs}
\usepackage{tipa}
\usepackage{amsfonts}
\usepackage{graphicx}
\usepackage{subfigure}
\usepackage[title]{appendix}
\usepackage{color}
\usepackage{multirow}
\usepackage{amsbsy}
\usepackage{longtable}
\usepackage{floatrow}
\usepackage{indentfirst}
\usepackage{cite}
\usepackage{paralist}
\allowdisplaybreaks[4] 
\newtheorem{theorem}{Theorem}
\newtheorem{definition}{Definition}

\newtheorem{lemma}{Lemma}

\begin{document}
\title {\bf Phaseless compressive sensing using partial support information}
\author{Zhiyong Zhou\footnote{Corresponding author, zhiyong.zhou@umu.se.}, Jun Yu\\
	Department of Mathematics and Mathematical Statistics, Ume{\aa} University, \\Ume{\aa},
	901 87, Sweden}
\maketitle
\date{}
\noindent
$\mathbf{Abstract}$: {We study the recovery conditions of weighted $\ell_1$ minimization for real-valued signal reconstruction from phaseless compressive sensing measurements when partial support information is available. A strong restricted isometry property condition is provided to ensure the stable recovery. Moreover, we present the weighted null space property as the sufficient and necessary condition for the success of $k$-sparse phaseless recovery via weighted $\ell_1$ minimization. Numerical experiments are conducted to illustrate our results.}
\\
$\mathbf{Keywords}$:  Phaseless compressive sensing; Partial support information; Strong restricted isometry property; Weighted null space property.
\section{Introduction}
Compressive sensing aims to recover an unknown signal from the underdetermined linear measurements (see \cite{ek,fr} for a comprehensive view). It is known as phase retrieval or phaseless compressive sensing when there is no phase information. The phaseless compressive sensing problem has recently attracted considerable research interests and many algorithms have been proposed to solve this problem. Existing literature include \cite{cesv,cls,csv,cc,gx,njs,sbe}, to name a few. Specifically, the goal of phaseless compressive sensing is to recover $x\in\mathbb{R}^N$ up to a unimodular scaling constant from noisy magnitude measurements $y=|Ax|+e\in\mathbb{R}^m$ with the measurement matrix  $A=(a_1,\cdots,a_m)^T\in\mathbb{R}^{m\times N}$, $|Ax|=(|\langle a_1,x\rangle|,\cdots,|\langle a_m,x\rangle|)^T$ and the noise term $e\in\mathbb{R}^m$. When $x$ is sparse or compressible, the stable recovery can be guaranteed by solving the following $\ell_1$ minimization problem
\begin{align}
\min\limits_{z\in\mathbb{R}^N}\,\lVert z\rVert_1\,\,\,\text{subject to\,\,\,$\lVert |Az|-y\rVert_2\leq \varepsilon$},
\end{align}
provided that the measurement matrix $A$ satisfies the strong restricted isometry property (SRIP) \cite{gwx,vx}. In the noiseless case, the first sufficient and necessary condition was presented in \cite{wx} by proposing a new version of null space property for the phase retrieval problem.

In this paper, we generalize the existing theoretical framework for phaseless compressive sensing to incorporate partial support information, where we consider the case that an estimate of the support of the signal is available. We follow the similar notations and arguments in \cite{fmsy,zy}. For an arbitrary signal $x\in\mathbb{R}^N$, let $x^k$ be its best $k$-term approximation, so that $x^k$ minimizes $\lVert x-f\rVert_{1}$ over all $k$-sparse vectors $f$. Let $T_0$ be the support of $x^k$, where $T_0\subset\{1,\cdots,N\}$ and $|T_0|\leq k$. Let $\tilde{T}$, the support estimate, be a subset of $\{1,2\cdots,N\}$ with cardinality $|\tilde{T}|=\rho k$, where $\rho\geq 0$ and $|\tilde{T}\cap T_0|=\alpha\rho k$ with $0\leq \alpha\leq 1$. Here the parameter $\rho$ determines the ratio of the size of the estimated support to the size of the actual support of $x^k$ (or the support of $x$ if $x$ is $k$-sparse), while the parameter $\alpha$ determines the ratio of the number of indices in the support of $x^k$ that are accurately estimated in $\tilde{T}$ to the size of $\tilde{T}$, i.e., $\alpha=\frac{|\tilde{T}\cap T_0|}{|\tilde{T}|}$. To incorporate prior support information $\tilde{T}$,  we adopt the weighted $\ell_1$ minimization \begin{align}
\min\limits_{z\in\mathbb{R}^N}\sum\limits_{i=1}^N \mathrm{w}_i |z_i|,\,\,\,\text{subject to $\lVert |Az|-y\rVert_2\leq\varepsilon$},\,\,\,
\text{where $\mathrm{w}_i=\begin{cases}
	\omega \in[0,1] &\text{$i \in\tilde{T}$,}\\
	1 &\text{$i\in\tilde{T}^c$.}
	\end{cases}$} \label{min}
\end{align}
We present the SRIP condition and weighted null space property condition to guarantee the success of the recovery via the weighted $\ell_1$ minimization problem above. 

The paper is organized as follows. In Section 2, we introduce the definition of SRIP and present the stable recovery condition with this tool. In Section 3, the sufficient and necessary weighted null space property condition for the real sparse noise free phase retrieval is given. In Section 4, some numerical experiments are presented to illustrate our theoretical results. Finally, Section 5 is devoted to the conclusion.

 Throughout the paper, for any vector $x\in\mathbb{R}^N$, we denote the $\ell_p$ norm by $\lVert x\rVert_p=(\sum_{i=1}^p |x_i|^p)^{1/p}$ for $p>0$ and the weighted $\ell_1$ norm as $\lVert x\rVert_{1,\mathrm{w}}=\sum_{i=1}^N \mathrm{w}_i |x_i|$. For any matrix $X$, $\lVert X\rVert_1$ denotes the entry-wise $\ell_1$ norm. For any set $T$, we denote its cardinality as $|T|$. The vector $x\in\mathbb{R}^N$ is called $k$-sparse if at most $k$ of its entries are nonzero, i.e., if $\lVert x\rVert_0=|\mathrm{supp}(x)|\leq k$, where $\mathrm{supp}(x)$ denotes the index set of the nonzero entries. We denote the index set $[N]:=\{1,2,\cdots,N\}$. For a matrix $A=(a_1,\cdots,a_m)^T\in\mathbb{R}^{m\times N}$ and an index set $I\subset[m]$, we denote $A_I$ the sub-matrix of $A$ where only rows with indices in $I$ are kept, i.e., $A_I=(a_j,j\in I)^T$. 

\section{SRIP}

To recover sparse signals via $\ell_1$ minimization in the classical compressive sensing setting, \cite{ct} introduced the notion of restricted isometry property (RIP) and established a sufficient condition. We say a matrix $A$ satisfies the RIP of order $k$ if there exists a constant $\delta_k\in[0,1)$ such that for all $k$-sparse vectors $x$ we have \begin{align}
(1-\delta_k)\lVert x\rVert_2^2\leq\lVert Ax\rVert_2^2\leq (1+\delta_k)\lVert x\rVert_2^2.
\end{align}
Cai and Zhang \cite{cz} proved that the RIP of order $tk$ with $\delta_{tk}<\sqrt{\frac{t-1}{t}}$ where $t>1$ can guarantee the exact recovery in the noiseless case and stable recovery in the noisy case via $\ell_1$ minimization. This condition is sharp when $t\geq \frac{4}{3}$, see \cite{cz} for details. Very recently, Chen and Li \cite{cl} generalized this sharp RIP condition to the weighted $\ell_1$ minimization problem when partial support information was incorporated. We first present the following useful lemma, which is an extension of the result in \cite{cl}.\\

\begin{lemma}
Let $x\in\mathbb{R}^N, y=Ax+e\in\mathbb{R}^m$ with $\lVert e\rVert_2\leq \zeta$, and $\eta\geq 0$. Suppose that $A$ satisfies RIP of order $tk$ with $\delta_{tk}<\sqrt{\frac{t-d}{t-d+\gamma^2}}$ for some $t>d$, where $\gamma=\omega+(1-\omega)\sqrt{1+\rho-2\alpha\rho}$ and \begin{align}
d=\begin{cases}
1, &\text{$\omega=1$}\\
1-\alpha\rho+a, &\text{$0\leq \omega<1$}
\end{cases}
\end{align}
with $a=\max\{\alpha,1-\alpha\}\rho$. Then for any $$
\hat{x}\in\{z\in\mathbb{R}^N:\lVert z\rVert_{1,\mathrm{w}}\leq\lVert x\rVert_{1,\mathrm{w}}+\eta,\lVert Az-y\rVert_2\leq \varepsilon\},
$$
we have \begin{align}
\lVert \hat{x}-x\rVert_2\leq C_1(\zeta+\varepsilon)+C_2\frac{2(\omega\lVert x_{T_0^c}\rVert_1+(1-\omega)\lVert x_{\tilde{T}^c\cap T_0^c}\rVert_1)}{\sqrt{k}}+C_2\frac{\eta}{\sqrt{k}},
\end{align}
where \begin{align*}
C_1&=\frac{\sqrt{2(t-d)(t-d+\gamma^2)(1+\delta_{tk})}}{(t-d+\gamma^2)(\sqrt{\frac{t-d}{t-d+\gamma^2}}-\delta_{tk})},\\
C_2&=\frac{\sqrt{2}\delta_{tk}\gamma+\sqrt{(t-d+\gamma^2)(\sqrt{\frac{t-d}{t-d+\gamma^2}}-\delta_{tk})\delta_{tk}}}{(t-d+\gamma^2)(\sqrt{\frac{t-d}{t-d+\gamma^2}}-\delta_{tk})}+\frac{1}{\sqrt{d}}.
\end{align*} \\
\end{lemma}

\noindent
{\bf Remark 1}\,\, Note that if $x^{\ell_2}$ is the solution of the weighted $\ell_1$ minimization problem: $$
\min\limits_{z\in\mathbb{R}^N}\,\,\lVert z\rVert_{1,\mathrm{w}},\,\,\text{subject to\,\,$\lVert Az-y\rVert_2\leq \varepsilon$},
$$ then $x^{\ell_2}\in\{z\in\mathbb{R}^N:\lVert z\rVert_{1,\mathrm{w}}\leq\lVert x\rVert_{1,\mathrm{w}}+\eta,\lVert Az-y\rVert_2\leq \varepsilon\}$ with $\eta=0$. Therefore, this lemma is an extension of Theorem 3.1 in \cite{cl} by letting $\zeta=\varepsilon$ and $\eta=0$. The proof follows from almost the same procedure for the proof of Theorem 3.1 in Section 4 of \cite{cl} via replacing the $P=\frac{2(\omega\lVert x_{T_0^c}\rVert_1+(1-\omega)\lVert x_{\tilde{T}^c\cap T_0^c}\rVert_1)}{\sqrt{k}\gamma}$ with $P'=\frac{2(\omega\lVert x_{T_0^c}\rVert_1+(1-\omega)\lVert x_{\tilde{T}^c\cap T_0^c}\rVert_1)+\eta}{\sqrt{k}\gamma}$, and letting $\zeta=\varepsilon$. In order not to repeat, we leave out all the details. In addition, this result also generalizes the Lemma 2.1 in \cite{gwx}, which is the special case with the noise term $e=0$, $\zeta=0$ and $\omega=1$. This lemma will play a crucial role in establishing the stable phaseless recovery result via weighted $\ell_1$ minimization later on.\\

To address the phaseless compressive sensing problem (\ref{min}), a stronger version of RIP is needed. Its definition is provided as follows.\\

\begin{definition} (SRIP \cite{gwx,vx})
We say a matrix $A=(a_1,\cdots,a_m)^T\in\mathbb{R}^{m\times N}$ has the Strong Restricted Isometry Property (SRIP) of order $k$ with bounds $\theta_{-},\theta_{+}\in(0,2)$ if \begin{align}
\theta_{-}\lVert x\rVert_2^2\leq \min\limits_{I\subseteq [m],|I|\geq m/2}\lVert A_{I}x\rVert_2^2\leq \max\limits_{I\subseteq [m],|I|\geq m/2}\lVert A_{I}x\rVert_2^2\leq \theta_{+}\lVert x\rVert_2^2 \label{srip}
\end{align}
holds for all $k$-sparse vectors $x\in\mathbb{R}^N$, where $[m]=\{1,\cdots,m\}$. We say $A$ has the Strong Lower Restricted Isometry Property of order $k$ with bound $\theta_{-}$ if the lower bound in (\ref{srip}) holds. Similarly, we say $A$ has the Strong Upper Restricted Isometry Property of order $k$ with bound $\theta_{+}$ if the upper bound in (\ref{srip}) holds.\\
\end{definition}

Next, we present the conditions for the stable recovery via weighted $\ell_1$ minimization by using SRIP.

\begin{theorem}
	Let $x\in\mathbb{R}^N, y=|Ax|+e\in\mathbb{R}^m$ with $\lVert e\rVert_2\leq \zeta$. Adopt the notations in Lemma 1 and assume that $A\in\mathbb{R}^{m\times N}$ satisfies the SRIP of order $tk$ with bounds $\theta_{-},\theta_{+}\in (0,2)$ such that \begin{align}
	t\geq \max\left\{d+\frac{\gamma^2(1-\theta_{-})^2}{2\theta_{-}-\theta_{-}^2},d+\frac{\gamma^2(1-\theta_{+})^2}{2\theta_{+}-\theta_{+}^2}\right\}.\label{sripc}
	\end{align}
Then any solution $x^{\sharp}$ of (\ref{min}) satisfies \begin{align}
\min\{\lVert x^{\sharp}-x\rVert_2,\lVert x^{\sharp}+x\rVert_2\}\leq C_1(\zeta+\varepsilon)+C_2\frac{2(\omega\lVert x_{T_0^c}\rVert_1+(1-\omega)\lVert x_{\tilde{T}^c\cap T_0^c}\rVert_1)}{\sqrt{k}}. \label{stable}
\end{align}
where $C_1$ and $C_2$ are constants defined in Lemma 1. \\
\end{theorem}

\noindent
{\bf Remark 2}\,\, As it has been proved in \cite{vx} that Gaussian matrices with $m=O(tk\log(N/k))$ satisfy SRIP of order $tk$ with high probability, thus the stable recovery result (\ref{stable}) can be achieved by using Gaussian measurement matrix with appropriate number of measurements $m$.\\

\noindent
{\bf Remark 3}\,\, Note that when the weight $\omega=1$, we have $\gamma=d=1$. Then, by assuming $\zeta=\varepsilon=0$ and $x$ is exactly $k$-sparse, our theorem reduces to Theorem 2.2 in \cite{vx}. That is, if $A$ satisfies the SRIP of order $tk$ with bounds $\theta_{-},\theta_{+}$ and $t\geq\max\{\frac{1}{2\theta_{-}-\theta_{-}^2},\frac{1}{2\theta_{+}-\theta_{+}^2}\}$, then for any $k$-sparse signal $x\in\mathbb{R}^N$ we have $\mathop{\arg\min}_{z\in\mathbb{R}^N}\{\lVert z\rVert_{1}: |Az|=|Ax|\}=\{\pm x\}$. Similarly, if we let the noise term $e=0$, $\zeta=0$ and $\omega=1$, this theorem goes to Theorem 3.1 in \cite{gwx}.\\

\noindent
{\bf Remark 4}\,\, If $\alpha=\frac{1}{2}$, we have $\gamma=d=1$. The sufficient condition (\ref{sripc}) of Theorem 1 is identical to that of Theorem 2.2 in \cite{vx} and that of Theorem 3.1 in \cite{gwx}. And the constants $C_1=c_1=\frac{\sqrt{2(1+\delta_{tk})}}{1-\sqrt{t/(t-1)}\delta_{tk}}, C_2=c_2=\frac{\sqrt{2}\delta_{tk}+\sqrt{(\sqrt{t(t-1)}-\delta_{tk}t)\delta_{tk}}}{\sqrt{t(t-1)-\delta_{tk}t}}$ (see Theorem 3.1 in \cite{gwx}). In addition, if $0\leq \omega<1$ and $\alpha>\frac{1}{2}$, then $d=1$ and $\gamma<1$. The sufficient condition (\ref{sripc}) in Theorem 1 is weaker than that of Theorem 2.2 in \cite{vx} and that of Theorem 3.1 in \cite{gwx}. In this case, the constants $C_1<c_1, C_2<c_2$.\\

 Set $t^{\omega}=\max\left\{d+\frac{\gamma^2(1-\theta_{-})^2}{2\theta_{-}-\theta_{-}^2},d+\frac{\gamma^2(1-\theta_{+})^2}{2\theta_{+}-\theta_{+}^2}\right\}$.
 We illustrate how the constants $t^{\omega}$, $C_1$ and $C_2$ change with $\omega$ for different values of $\alpha$ in Figure 1. In all the plots, we set $\rho=1$. In the plot of $t^{\omega}$, we set $\theta_{-}=\frac{1}{2}$ and $\theta_{+}=\frac{3}{2}$, then $t^{\omega}=d+\frac{\gamma^2}{3}$. In the plots of $C_1$ and $C_2$, we fix $t=4$ and $\delta_{tk}=0.3$. Note that if $\omega=1$ or $\alpha=0.5$, then $t^{\omega}\equiv 1+\frac{1}{3}=\frac{4}{3}$, $C_1\equiv c_1$ and $C_1\equiv c_2$. And it shows that $t^{\omega}$ decreases as $\alpha$ increases, which means that the sufficient condition (\ref{sripc}) becomes weaker as $\alpha$ increases. For each $\alpha>0.5$, the sufficient condition becomes stronger ($t^{\omega}$ increases) as $\omega$ increases. For instance, if $90\%$ of the support estimate is accurate ($\alpha=0.9$) and $\omega=0.6$, we have $t^{\omega}=1.2022$, while $t^{\omega}=1.3333$ for standard $\ell_1$ minimization ($\omega=1$). The opposite conclusion holds for the case $\alpha<0.5$. In addition, as $\alpha$ increases, the constant $C_1$ decreases with $t=4$ and $\delta_{tk}=0.3$. Meanwhile, the constant $C_2$ with $\alpha\neq 0.5$ is smaller than that with $\alpha=0.5$.  \\
 
 \begin{figure}[htbp]
 	\centering
 	\includegraphics[width=\textwidth,height=0.4\textheight]{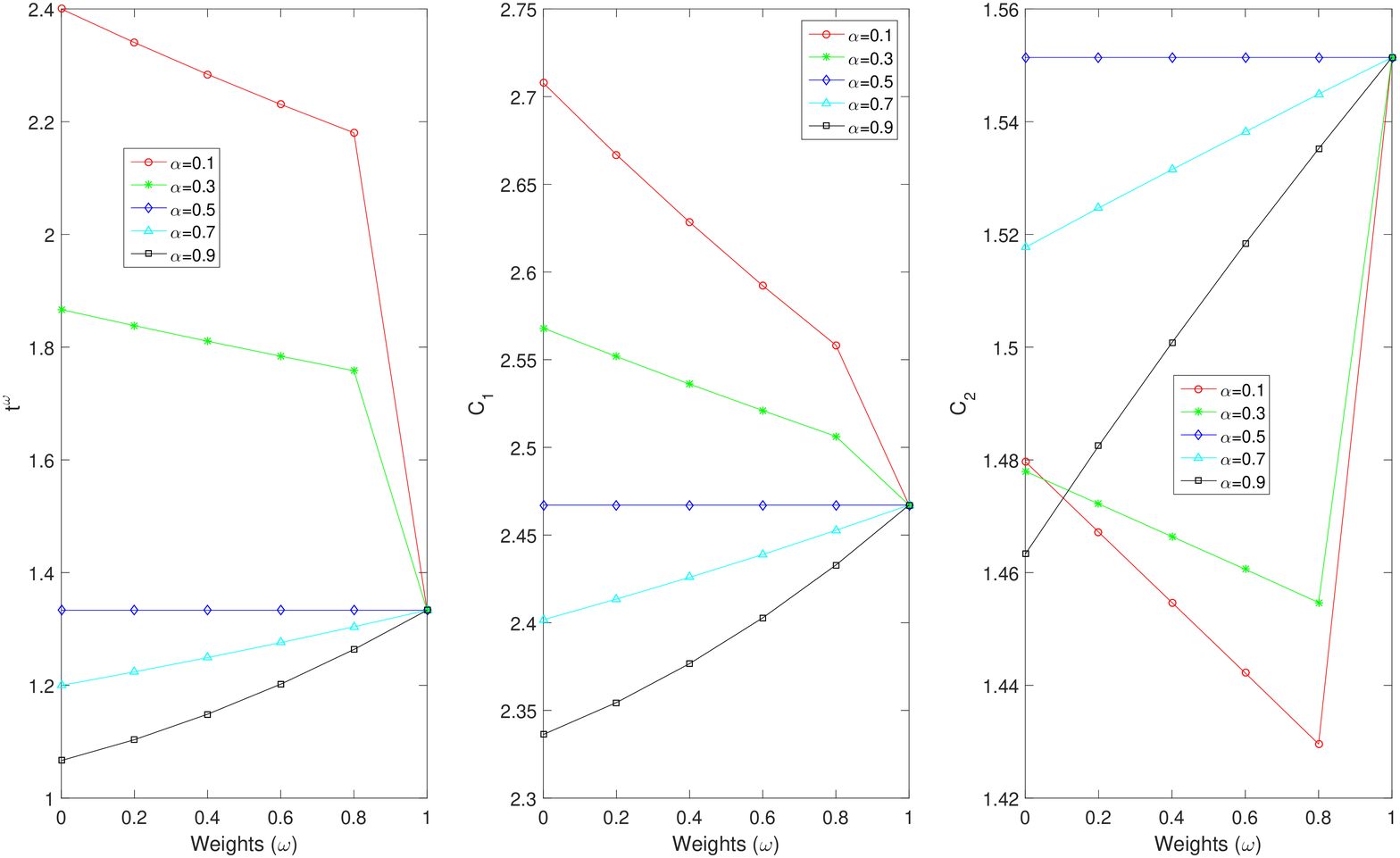}
 	\caption{Comparison of the constants $t^{\omega}$, $C_1$ and $C_2$ for various of $\alpha$. In all the plots, we set $\rho=1$. In the plot of $t^{\omega}$, we set $\theta_{-}=\frac{1}{2}$ and $\theta_{+}=\frac{3}{2}$. In the plots of $C_1$ and $C_2$, we fix $t=4$ and $\delta_{tk}=0.3$.}\label{fig:1}
 \end{figure}

\noindent
{\bf Proof of Theorem 1.} For any solution $x^{\sharp}$ of (\ref{min}), we have $$
\lVert x^{\sharp}\rVert_{1,\mathrm{w}}\leq \lVert x\rVert_{1,\mathrm{w}}
$$
and $$
\lVert |Ax^{\sharp}|-|Ax|-e\rVert_2\leq \varepsilon.
$$

If we divide the index set $\{1,2,\cdots,m\}$ into two subsets \begin{align*}
T=\{j: \mathrm{sign}(\langle a_j,x^{\sharp}\rangle)=\mathrm{sign}(\langle a_j,x\rangle)\}\,\,\,\text{and}\,\,\,T^c=\{j: \mathrm{sign}(\langle a_j,x^{\sharp}\rangle)=-\mathrm{sign}(\langle a_j,x\rangle)\},
\end{align*}
then it implies that \begin{align}
\lVert A_T x^{\sharp}-A_T x-e\rVert_2+\lVert A_{T^c} x^{\sharp}+A_{T^c} x-e\rVert_2\leq \varepsilon.
\end{align}
Here either $|T|\geq m/2$ or $|T^c|\geq m/2$. If $|T|\geq m/2$, we use the fact that \begin{align}
\lVert A_T x^{\sharp}-A_T x-e\rVert_2\leq \varepsilon.
\end{align}
Then, we obtain $$
x^{\sharp}\in\{z\in\mathbb{R}^N:\lVert z\rVert_{1,\mathrm{w}}\leq \lVert x\rVert_{1,\mathrm{w}},\lVert A_T z-A_T x-e\rVert_2\leq \varepsilon\}.
$$
Since $A$ satisfies SRIP of order $tk$ with bounds $\theta_{-},\theta_{+}$ and $$
t\geq \max\left\{d+\frac{\gamma^2(1-\theta_{-})^2}{2\theta_{-}-\theta_{-}^2},d+\frac{\gamma^2(1-\theta_{+})^2}{2\theta_{+}-\theta_{+}^2}\right\}>d,
$$
therefore, the definition of SRIP implies that $A_T$ satisfies the RIP of order $tk$ with \begin{align}
\delta_{tk}\leq\max\{1-\theta_{-},\theta_{+}-1\}\leq \sqrt{\frac{t-d}{t-d+\gamma^2}}.
\end{align}

Thus, by using Lemma 1 with $\eta=0$, we have $$
\lVert x^{\sharp}-x\rVert_2\leq C_1(\zeta+\varepsilon)+C_2\frac{2(\omega\lVert x_{T_0^c}\rVert_1+(1-\omega)\lVert x_{\tilde{T}^c\cap T_0^c}\rVert_1)}{\sqrt{k}}.
$$
Similarly, if $|T^c|\geq m/2$, we obtain the other corresponding result $$
\lVert x^{\sharp}+x\rVert_2\leq C_1(\zeta+\varepsilon)+C_2\frac{2(\omega\lVert x_{T_0^c}\rVert_1+(1-\omega)\lVert x_{\tilde{T}^c\cap T_0^c}\rVert_1)}{\sqrt{k}}.
$$
The proof of Theorem 1 is now completed.

\section{Weighted Null Space Property}
In this section, we consider the noiseless weighted $\ell_1$ minimization problem, i.e., \begin{align}
\min\limits_{z\in\mathbb{R}^N}\,\,\lVert z\rVert_{1,\mathrm{w}},\,\,\,\text{subject to $|Az|=|Ax|$},\,\,\,
\text{where $\mathrm{w}_i=\begin{cases}
	\omega \in[0,1], &\text{$i \in\tilde{T}$}\\
	1, &\text{$i\in\tilde{T}^c$}
	\end{cases}$}.
\end{align}
We denote the kernel space of $A$ by $\mathcal{N}(A):=\{h\in\mathbb{R}^N: Ah=0\}$ and denote the $k$-sparse vector space $\Sigma_{k}^N:=\{x\in\mathbb{R}^N:\lVert x\rVert_0\leq k\}$.\\

\begin{definition}
	 The matrix $A$ satisfies the $\mathrm{w}$-weighted null space property of order $k$ if for any nonzero $h\in \mathcal{N}(A)$ and any $T\subset[N]$ with $|T|\leq k$ it holds that \begin{align}
	\lVert h_T\rVert_{1,\mathrm{w}}<\lVert h_{T^c}\rVert_{1,\mathrm{w}}, \label{nsp}
	\end{align}
	where $T^c$ is the complementary index set of $T$ and $h_T$ is the restriction of $h$ to $T$. \\
\end{definition}

\noindent
{\bf Remark 5}\,\, Obviously, when the weight $\omega=1$, the weighted null space property reduces to the classical null space property. And according to the specific setting of $\mathrm{w}_i$, the expression (\ref{nsp}) is equivalent to $$
\omega\lVert h_{T\cap\tilde{T}}\rVert_1+\lVert h_{T\cap \tilde{T}^c}\rVert_1<\omega\lVert h_{T^c\cap\tilde{T}}\rVert_1+\lVert h_{T^c\cap \tilde{T}^c}\rVert_1\Leftrightarrow \omega\lVert h_T\rVert_1+(1-\omega)\lVert h_G\rVert_1<\lVert h_{T^c}\rVert_1,
$$
where $G=(T\cap\tilde{T}^c)\cup(T^c\cap \tilde{T})$ (see \cite{ms} for more arguments).\\

It is known that a signal $x\in\Sigma_k^N$ can be recovered via the weighted $\ell_1$ minimization problem if and only if the measurement matrix $A$ has the weighted null space property of order $k$. We state it as follows (see \cite{zxwk}):\\

\begin{lemma}
Given $A\in\mathbb{R}^{m\times N}$, for every $k$-sparse vector $x\in\mathbb{R}^N$ it holds that $$\mathop{\arg\min}\limits_{z\in\mathbb{R}^N}\,\,\{\lVert z\rVert_{1,\mathrm{w}}: Az=Ax\}=x
$$ if and only if $A$ satisfies the $\mathrm{w}$-weighted null space property of order $k$. \\
\end{lemma}

Next, we extend Lemma 2 to the following theorem on phaseless compressive sensing for the real-valued signal reconstruction.\\

\begin{theorem}
The following statements are equivalent:\\
(a) For any $k$-sparse $x\in\mathbb{R}^N$, we have \begin{align}
\mathop{\arg\min}\limits_{z\in\mathbb{R}^N}\{\lVert z\rVert_{1,\mathrm{w}}:|Az|=|Ax|\}=\{\pm x\}.
\end{align}\\
(b) For every $S\subseteq [m]$, it holds\begin{align}
\lVert u+v\rVert_{1,\mathrm{w}}<\lVert u-v\rVert_{1,\mathrm{w}} \label{pnsp}
\end{align}
for all nonzero $u\in \mathcal{N}(A_S)$ and $v\in \mathcal{N}(A_{S^c})$ satisfying $\lVert u+v\rVert_0\leq k$.\\
\end{theorem}

\noindent
{\bf Remark 6}\,\, If $\omega=1$, then Theorem 2 reduces to Theorem 3.2 in \cite {wx}. Since $\mathrm{w}_i=\omega$ when $i\in\tilde{T}$, and $\mathrm{w}_i=1$ otherwise, the expression (\ref{pnsp}) is equivalent to \begin{align*}
\omega\lVert u+v\rVert_1+(1-\omega)\lVert (u+v)_{\tilde{T}^c}\rVert_1<\omega\lVert u-v\rVert_1+(1-\omega)\lVert (u-v)_{\tilde{T}^c}\rVert_1.
\end{align*}
\\
\noindent
{\bf Proof of Theorem 2.} The proof follows from the proof of Theorem 3.2 in \cite{wx} with minor modifications. First we show $(a)\Rightarrow(b)$. Assume (b) is false, that is, there exist nonzero $u\in \mathcal{N}(A_S)$ and $v\in \mathcal{N}(A_{S^c})$ such that $$
\lVert u+v\rVert_{1,\mathrm{w}}\geq\lVert u-v\rVert_{1,\mathrm{w}}
$$
and $u+v\in\Sigma_{k}^N$. Now set $x=u+v\in\Sigma_k^N$, obviously for $i=1,\cdots,m$, we have $$
|\langle a_i, x\rangle|=|\langle a_i,u+v\rangle|=|\langle a_i,u-v\rangle|,
$$
since either $\langle a_i,u\rangle=0$ or $\langle a_i,v\rangle=0$. In other words $|Ax|=|A(u-v)|$. Note that $u-v\neq-x$, for otherwise we would have $u=0$, which is a contradiction. Then, it follows from (a) that we obtain \begin{align*}
\lVert x\rVert_{1,\mathrm{w}}=\lVert u+v\rVert_{1,\mathrm{w}}<\lVert u-v\rVert_{1,\mathrm{w}},
\end{align*}
This is a contradiction. Thus, (b) holds.\\ 

Next we prove $(b)\Rightarrow (a)$. Let $b=(b_1,\cdots,b_m)^T=|Ax|$ where $x\in\Sigma_k^N$. For a fixed $\sigma=(\sigma_1,\cdots,\sigma_m)^T\in\{-1,1\}^m$, we set $b^{\sigma}=(\sigma_1b_1,\cdots,\sigma_m b_m)^T$. We now consider the following weighted $\ell_1$ minimization problem: \begin{align}
\min\limits_{z\in\mathbb{R}^N}\,\,\lVert z\rVert_{1,\mathrm{w}}\,\,\,\text{subject to\,\,\,$Az=b^{\sigma}$}.
\end{align}
Its solution is denoted as $x^{\sigma}$. Then, we claim that for any $\sigma\in\{1,-1\}^m$, if $x^{\sigma}$ exists (it may not exist), we have $$
\lVert x^{\sigma}\rVert_{1,\mathrm{w}}\geq \lVert x\rVert_{1,\mathrm{w}}
$$ and the equality holds if and only if $x^{\sigma}=\pm x$. 

To prove the claim, we assume $\sigma^{\star}\in\{1,-1\}^m$ such that $b^{\sigma^{\star}}=Ax$. First note that the statement (b) implies the classical weighted null space property of order $k$. To see this, for any nonzero $h\in \mathcal{N}(A)$ and $T\subseteq [N]$ with $|T|\leq k$, we set $u=h$, $v=h_T-h_{T^c}$ and $S=[m]$. Then, we have $u\in \mathcal{N}(A_S)$ and $v\in \mathcal{N}(A_{S^c})$. Therefore, the statement (b) now implies \begin{align*}
2\lVert h_T\rVert_{1,\mathrm{w}}=\lVert u+v\rVert_{1,\mathrm{w}}<\lVert u-v\rVert_{1,\mathrm{w}}=2\lVert h_{T^c}\rVert_{1,\mathrm{w}}.
\end{align*}
As a consequence, we have $x^{\sigma^{\star}}=x$ by Lemma 2. And, similarly we have $x^{-\sigma^{\star}}=-x$. Next, for any $\sigma\in\{-1,1\}^m\neq \pm \sigma^{\star}$, if $x^{\sigma}$ doesn't exist then we have nothing to prove. Assume it does exist, set $S_{\star}=\{i: \sigma_i=\sigma^{\star}_i\}$. Then \begin{align*}
\langle a_i, x^{\sigma}\rangle=\begin{cases}
\langle a_i,x\rangle &i\in S_{\star},\\
-\langle a_i,x\rangle &i\in S_{\star}^c.
\end{cases}
\end{align*}
Set $u=x-x^{\sigma}$ and $v=x+x^{\sigma}$. Obviously, $u\in \mathcal{N}(A_{S_{\star}})$ and $v\in \mathcal{N}(A_{S_{\star}^c})$. Furthermore, $u+v=2x\in\Sigma_k^N$. Then, by the statement (b), we have $$
2\lVert x\rVert_{1,\mathrm{w}}=\lVert u+v\rVert_{1,\mathrm{w}}<\lVert u-v\rVert_{1,\mathrm{w}}=2\lVert x^{\sigma}\rVert_{1,\mathrm{w}}.
$$
This proves (a) and the proof is completed. 

\section{Simulations}
In this section, we present some simple numerical experiments to illustrate the benefits of using weighted $\ell_1$ minimization to recover sparse and compressible signals when partial prior support information is available in the phaseless compressive sensing case. In order to facilitate the computation, we follow a non-standard noise model: \begin{align}
b=|Ax|^2+e=\{a_i^T xx^Ta_i\}_{1\leq i\leq m}+e,
\end{align}
where $e\in\mathbb{R}^m$ is a noise term with $\lVert e\rVert_2\leq \varepsilon$. Then the weighted $\ell_1$ minimization goes to \begin{align}
\min\limits_{z\in\mathbb{R}^N}\sum\limits_{i=1}^N \mathrm{w}_i |z_i|,\,\,\,\text{subject to $\lVert |Az|^2-b\rVert_2\leq\varepsilon$},\,\,\,
\text{where $\mathrm{w}_i=\begin{cases}
	\omega \in[0,1] &\text{$i \in\tilde{T}$,}\\
	1 &\text{$i\in\tilde{T}^c$.}
	\end{cases}$}  \label{phaseless}
\end{align}

Here we adopt the compressive phase retrieval via lifting (CPRL) algorithm developed in \cite{oyds} to solve this phaseless recovery problem. 
By using a lifting technique, this problem can be rewritten as a semidefinite program (SDP). More specifically, given the ground truth signal $x\in\mathbb{R}^N$, let $X=xx^{T}\in\mathbb{R}^{N\times N}$ be an induced rank-1 semidefinite matrix. We further denote $\Phi_i=a_i a_i^T$, a linear operator $B$ of $Z=zz^{T}\in\mathbb{R}^{N\times N}$ as \begin{align*}
B: Z\mapsto \{\mathrm{Tr}(\Phi_i Z)\}_{1\leq i\leq m}\in\mathbb{R}^m
\end{align*} and the weight matrix $W=\mathrm{diag}\{\mathrm{w}_i, 1\leq i\leq N\}\in\mathbb{R}^{N\times N}$. Then the phaseless vector recovery problem (\ref{phaseless}) can be cast as the following rank-1 matrix recovery problem: \begin{align*}
\min_{Z\in\mathbb{R}^{N\times N}}\,\,&\lVert WZW^{T}\rVert_1, \\ 
\text{subject to}\,\,\, &\lVert B(Z)-b\rVert_2\leq \varepsilon,\\
&\mathrm{rank}(WZW^{T})=1, Z\succeq 0.
\end{align*}
This is of course still a non-convex problem due to the rank constraint. The lifting approach addresses this issue by replacing $\mathrm{rank}(WZW^{T})$ with $\mathrm{Tr}(WZW^{T})$. This leads to an SDP: 
\begin{align}
\min_{Z\in\mathbb{R}^{N\times N}}\,\,&\mathrm{Tr}(WZW^{T})+\lambda\lVert WZW^{T}\rVert_1, \nonumber \\ 
\text{subject to}\,\,\, &\lVert B(Z)-b\rVert_2\leq \varepsilon,   \nonumber \\
 &Z\succeq 0, \label{sdp}
\end{align}
where $\lambda>0$ is a design parameter. Then the estimate of $x$ can be finally be found by computing the rank-1 decomposition of the recovered matrix via singular value decomposition. 

The recovery performance is assessed by the average reconstruction signal to noise ratio (SNR) over 10 experiments. The SNR is measured in dB and it is given by \begin{align}
\mathrm{SNR}(x,x^{\sharp})=20\log_{10}\left(\frac{\lVert x\rVert_2}{\min\{\lVert x^{\sharp}-x\rVert_2,\lVert x^{\sharp}+x\rVert_2\}}\right),
\end{align} 
where $x$ is the true signal and $x^{\sharp}$ is the recovered signal. For all the experiments, we fix the parameter $\lambda=1$. In the experiments where the measurements are noisy, we set the noise $\{e_i,1\leq i\leq m\}\overset{i.i.d}\sim N(0,\sigma^2)$ with $\sigma=0.1$ and $\varepsilon=\lVert e\rVert_2$.

\subsection{Sparse Case}
We first consider the case that $x$ is exactly sparse with an ambient dimension $N=32$ and fixed sparsity $k=4$. The sparse signals are generated by choosing $k$ nonzero positions uniformly at random, and then choosing the nonzero values from the standard normal distribution for these k nonzero positions. The recovery is done via (\ref{sdp}) using a support estimate of size $|\tilde{T}|=4$ (i.e., $\rho=1$).

Figure 2 shows the recovery performances for different $\alpha$ and $\omega$ with an increasing number of measurements $m$, both in the noise free and noisy cases. It can be observed that when $\alpha=0.75>0.5$, the best recovery is achieved for very small $\omega$ whereas a $\omega=1$ results in the lowest SNR for both cases. On the other hand, when $\alpha=0.25< 0.5$, the performance of the recovery algorithms is better for large $\omega$ than that for small $\omega$. The case $\omega=0$ results in the lowest SNR. When $\alpha=0.5$, the performance gaps for different $\omega$ are not particularly large and it seems that a medium $\omega$ ($\omega=0.5$) achieves the best recovery. In the noise free case, a perfect recovery can be achieved as long as the number of measurements $m$ is large enough. As is also expected that in all settings, comparing to the noise free case, we have a lower SNR in the noisy case. These findings are largely consistent with the theoretical results provided in Section 2.

\begin{figure}[htbp]
	\centering
	\includegraphics[width=\textwidth,height=0.4\textheight]{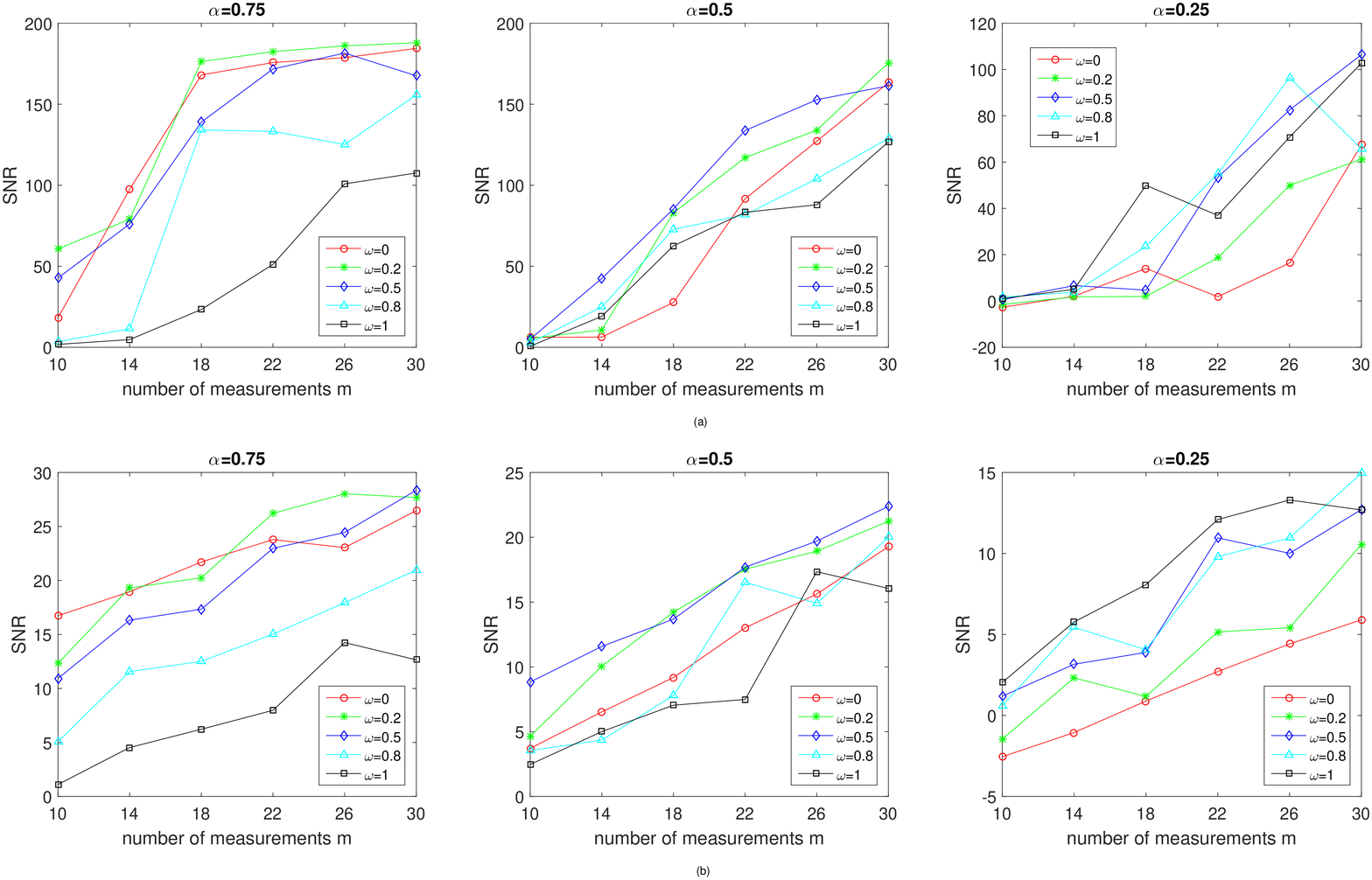}
	\caption{Performance of weighted $\ell_1$ recovery in terms of SNR averaged over 10 experiments for sparse signals $x$ with $N=32$, $k=4$, while varying the number of measurements $m$. From left to right, $\alpha=0.75$, $\alpha=0.5$ and $\alpha=0.25$. (a) Noise Free. (b) $\sigma=0.1$.}\label{fig:2}
\end{figure}
\subsection{Compressible Case}

Here we generate a signal $x$ whose coefficients decay like $j^{-\theta}$ where $j\in\{1,\cdots,N\}$ and $\theta=4.5$. This kind of signal itself is not sparse, but can be well approximated by an exactly sparse signal. For this experiment, we set $k=4$, i.e., we use the best 4-term approximation. We fix $\rho=1$ as in the sparse case. The phaseless recovery results are presented in Figure 3. It shows that on average a mediate value of $\omega$ ($\omega=0.5$) results in the best recovery. In general, when $\alpha>0.5$, smaller $\omega$ favours better reconstruction results. The opposite conclusion holds for the case that $\alpha<0.5$. Therefore, as is expected that the behaviors that occur in the exactly sparse case also occur in the compressible case. 

\begin{figure}[htbp]
	\centering
	\includegraphics[width=\textwidth,height=0.4\textheight]{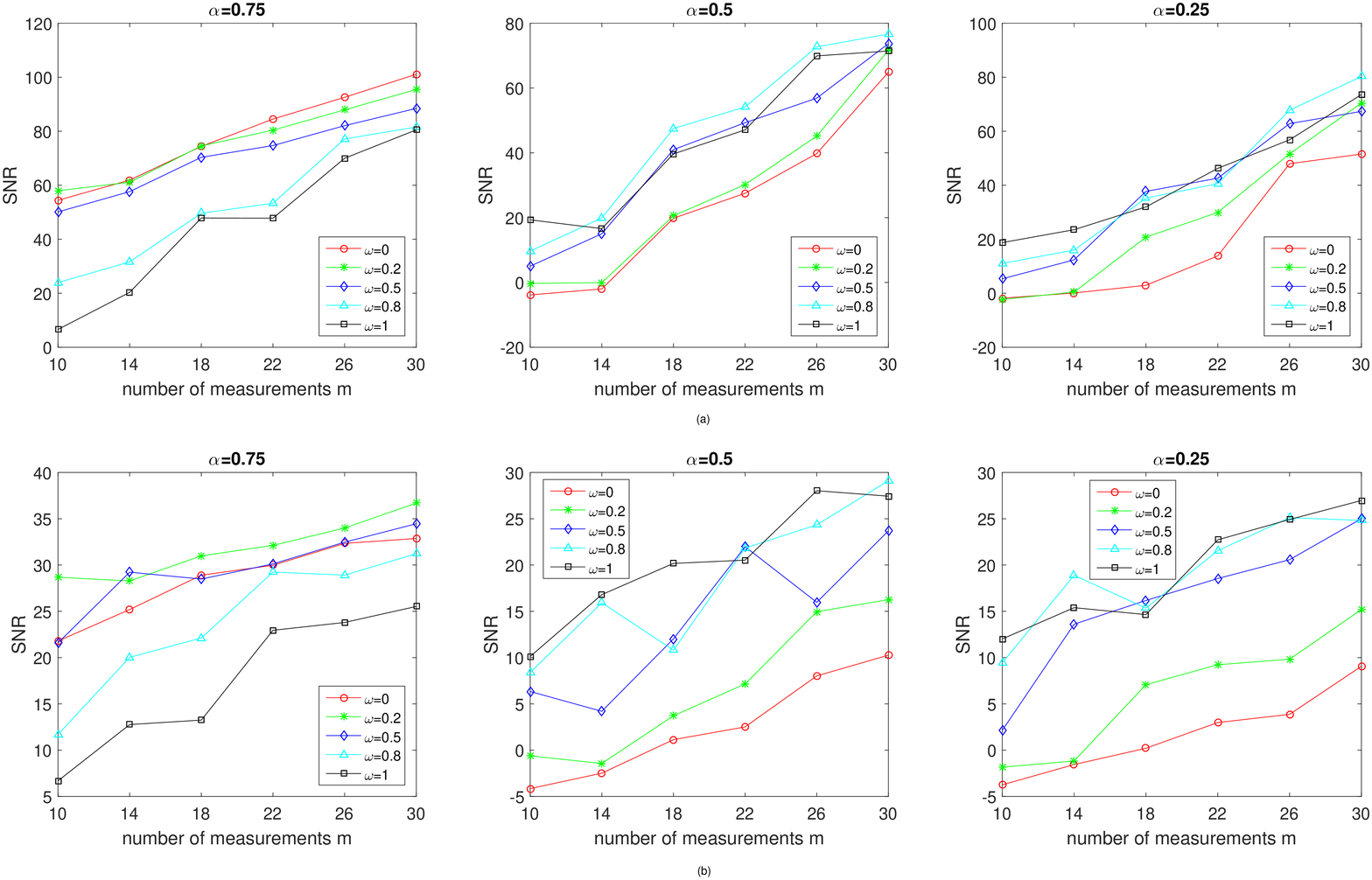}
	\caption{Performance of weighted $\ell_1$ recovery in terms of SNR averaged over 10 experiments for compressible signals $x$ with $N=32$, $\theta=4.5$, while varying the number of measurements $m$. From left to right, $\alpha=0.75$, $\alpha=0.5$ and $\alpha=0.25$. (a) Noise Free. (b) $\sigma=0.1$.}\label{fig:3}
\end{figure}

\section{Conclusion}

In this paper, we established the sufficient SRIP condition and the sufficient and necessary weighted null space property condition for phaseless compressive sensing using partial support information via weighted $\ell_1$ minimization, and we conducted some numerical experiments to illustrate the theoretical results. 

Some further problems are left for future work. As we only consider the real-valued signal reconstruction case, it will be challenging to generalize the present results to the complex-valued signal case. Besides it will be very interesting to construct the measurement matrix $A\in\mathbb{R}^{m\times N}$ satisfying the weighted null space property given in (\ref{pnsp}) directly.

\section*{Acknowledgements}
This work is supported by the Swedish Research Council grant (Reg.No. 340-2013-5342).


\begin{thebibliography}{}

\bibitem{cz}	
Cai, T. T. and Zhang, A. (2014). Sparse representation of a polytope and recovery of sparse signals and low-rank matrices. \textit{IEEE Transactions on Information Theory} \textbf{60}(1) 122--132.

\bibitem{cesv}
Candes, E. J., Eldar, Y. C., Strohmer, T. and Voroninski, V. (2015). Phase retrieval via matrix completion. \textit{SIAM review}
\textbf{57}(2) 225--251.

\bibitem{cls}
Candes, E. J., Li, X. and Soltanolkotabi, M. (2015). Phase retrieval via Wirtinger flow: Theory and algorithms. \textit{IEEE Transactions on Information Theory}
\textbf{61}(4) 1985--2007.

\bibitem{csv}
Candes, E. J., Strohmer, T. and Voroninski, V. (2013). Phaselift: Exact and stable signal recovery from magnitude measurements via convex programming. \textit{Communications on Pure and Applied Mathematics} 
\textbf{66}(8) 1241--1274.
	
\bibitem{ct}
Candes, E. J. and Tao, T. (2005). Decoding by linear programming. \textit{IEEE Transactions on Information Theory}  \textbf{51}(12) 4203--4215.

\bibitem{cl}
Chen, W. and Li, Y. (2016). Recovery of signals under the high order RIP condition via prior support information. \textit{arXiv preprint arXiv:1603.03464.}

\bibitem{cc}
Chen, Y. and Candes, E. (2015). Solving random quadratic systems of equations is nearly as easy as solving linear systems. In \textit{Advances in Neural Information Processing Systems} (pp. 739--747).

\bibitem{ek}
Eldar, Y. C. and Kutyniok, G. (2012). Compressed Sensing: Theory and Applications. Cambridge University Press.

\bibitem{fr}
Foucart, S. and Rauhut, H. (2013). A Mathematical Introduction to Compressive Sensing. New York, NY, USA: Springer-Verlag.

\bibitem{fmsy}
Friedlander, M. P., Mansour, H., Saab, R. and Yilmaz, O. (2012). Recovering compressively sampled signals using partial support information. \textit{IEEE Transactions on Information Theory}
\textbf{58}(2) 1122--1134.

\bibitem{gwx}
Gao, B., Wang, Y. and Xu, Z. (2016). Stable signal recovery from phaseless measurements. \textit{Journal of Fourier Analysis and Applications}
\textbf{22}(4) 787--808.

\bibitem{gx}
Gao, B. and Xu, Z. (2017). Phaseless recovery using the Gauss–Newton method. \textit{IEEE Transactions on Signal Processing}
\textbf{65}(22) 5885--5896.

\bibitem{ms}
Mansour, H. and Saab, R. (2017). Recovery analysis for weighted $\ell_1$-minimization using the null space property. \textit{Applied and Computational Harmonic Analysis} \textbf{43}(1) 23--38.

\bibitem{njs}
Netrapalli, P., Jain, P. and Sanghavi, S. (2015). Phase retrieval using alternating minimization. \textit{IEEE Transactions on Signal Processing}
\textbf{18}(63) 4814--4826.

\bibitem{oyds}
Ohlsson, H., Yang, A. Y., Dong, R. and Sastry, S. S. (2011). Compressive phase retrieval from squared output measurements via semidefinite programming. \textit{arXiv preprint arXiv:1111.6323.}

\bibitem{sbe}
Shechtman, Y., Beck, A. and Eldar, Y. C. (2014). GESPAR: Efficient phase retrieval of sparse signals. \textit{IEEE Transactions on Signal Processing}
\textbf{62}(4) 928--938.

\bibitem{vx}
Voroninski, V. and Xu, Z. (2016). A strong restricted isometry property, with an application to phaseless compressed sensing. \textit{Applied and Computational Harmonic Analysis}
\textbf{40}(2) 386--395.

\bibitem{wx}
Wang, Y. and Xu, Z. (2014). Phase retrieval for sparse signals. \textit{Applied and Computational Harmonic Analysis}
\textbf{37}(3) 531--544.

\bibitem{zxwk}
Zhou, S., Xiu, N., Wang, Y. and Kong, L. (2013). Exact recovery for sparse signal via weighted $\ell_1$ minimization. \textit{arXiv preprint arXiv:1312.2358.}

\bibitem{zy}
Zhou, Z. and Yu, J. (2017). Recovery analysis for weighted mixed $\ell_2/\ell_p$ minimization with $0<p\leq 1$. \textit{arXiv preprint arXiv:1709.00257.}
\end{thebibliography}
\end{document}